# Tunable Quasiparticle Band Gap in Few Layer GaSe/graphene Van der Waals Heterostructures

Zeineb Ben Aziza[1], Debora Pierucci[1], Hugo Henck[1], Mathieu G. Silly[2], Christophe David[1], Mina Yoon[3, 4], Fausto Sirotti[2], Kai Xiao[3], Mahmoud Eddrief[5, 6], Jean-Christophe Girard[1], and Abdelkarim Ouerghi[1,*]

[1] Centre de Nanosciences et de Nanotechnologies, CNRS, Univ. Paris-Sud, Université Paris-Saclay, C2N – Marcoussis, 91460 Marcoussis, France
[2] Synchrotron-SOLEIL, Saint-Aubin, BP48, F91192 Gif sur Yvette Cedex, France
[3] Center for Nanophase Materials Sciences, Oak Ridge National Laboratory, Oak Ridge, Tennessee 37831, USA
[4] The University of Tennessee, Knoxville, Tennessee 37996, USA
[5] Sorbonne Universités, UPMC Univ. Paris 06, UMR 7588, INSP, F-75005 Paris, France
[6] CNRS, UMR 7588, Institut des NanoSciences de Paris (INSP), F-75005 Paris, France

*Corresponding author, E-mail: abdelkarim.ouerghi@lpn.cnrs.fr, fax number: +33169636006

Two-dimensional (2D) materials have recently been the focus of extensive research. By following a similar trend as graphene, other 2D materials including transition metal dichalcogenides ($MX_2$) and metal mono-chalcogenides (MX) show great potential for ultrathin nanoelectronic and optoelectronic devices. Despite the weak nature of interlayer forces in semiconducting MX materials, their electronic properties are highly dependent on the number of layers. Using scanning tunneling microscopy and spectroscopy (STM/STS), we demonstrate the tunability of the quasiparticle energy gap of few layered gallium selenide (GaSe) directly grown on a bilayer graphene substrate by molecular beam epitaxy (MBE). Our results show that the band gap is about 3.50 ± 0.05 eV for single-tetralayer (1TL), 3.00 ± 0.05 eV for bi-tetralayer (2TL) and 2.30 ± 0.05 eV for tri-tetralayer (3TL). This band gap evolution of GaSe, in particularly the shift of the valence band with respect to the Fermi level, was confirmed by angle-resolved photoemission spectroscopy (ARPES) measurements and our theoretical calculations. Moreover, we observed a charge transfer in GaSe/graphene van der Waals (vdW) heterostructure using ARPES. These findings demonstrate the high impact on the GaSe electronic band structure and electronic properties that can be obtained by the control of 2D materials layer thickness and the graphene induced doping.

# I. INTRODUCTION

Graphene is a single layer of hexagonally ordered carbon atoms with super high carrier mobility[1]. However, the absence of the band gap restricts its application in the area of optoelectronics[2]. Thus, a lot of efforts have been devoted to open a band gap and improve the properties of graphene for such applications. Alternatively, on the new frontiers of 2D materials research, various other layered materials such as hexagonal boron nitride (h-BN)[3], transition metal dichalcogenides ($MX_2$) and the less known family of semiconducting metal monochalcogenides (MX) have been recently synthesized or isolated[2].

Gallium selenide (GaSe) is a layered MX crystal widely used in the field of optoelectronics[4,5], nonlinear optics[6], and terahertz experiments[7]. Recently, high photoresponse and on/off ratio of mechanically exfoliated GaSe based photodetectors have been reported[8]. However, for device fabrication, large areas of few TLs are required. Hence, the epitaxial growth of GaSe films seems to be a suitable synthesis method to promote the use of GaSe[9,10]. More interestingly, the possibility of combining semiconducting MX materials with graphene layer can merge the excellent optical properties of these materials with the high mobility of graphene in vdW heterostructures[11,12]. This combination is expected to give rise to novel materials capable of highly responsive photodetectors.

$MX_2$ and MX materials present thickness dependent properties[13]. For example in several semiconducting $MX_2$ materials (such as $MoS_2$, $WS_2$, $MoSe_2$, $WSe_2$...), a transition from an indirect band gap in bulk to a direct gap in single layer due to the quantum confinement was predicted and recently observed[14]. Unlike $MX_2$, the MX materials, especially GaSe, are predicted to present an indirect band gap in single layer structure, which evolves into a direct gap starting from roughly seven layers. First-principles calculations[15] predicted that 2D GaSe films (below 7 layers) have unusual band structures with the double valence band maximum (VBM), which could result in intriguing features such as van Hove singularity near the VBM for possible ferromagnetism[16,17]. While the electronic band gaps for single, bi, and tri TL GaSe have not yet been experimentally determined, an optical band gap of 2.1 eV has been measured for bulk GaSe using photoluminescence[18]. However, no experiment so far has directly confirmed the theoretical prediction of the unusual band structure and their evolution in terms of GaSe thickness[19,20]. These properties are one of the most promising properties of MX materials for device applications and are the focus of this work.

For this purpose we have opted for scanning tunneling microscopy/spectroscopy (STM/STS). STM/STS is a known powerful method for probing electronic structures of thin films and its studies have already been implemented to extract the quasi-particle band gaps and band edges in few layers $MX_2$[21,22]. In this work, we demonstrate experimentally the evolution of the band gap of GaSe as a function of the GaSe thickness. We reveal how the electronic band gap, the local density of states (LDOS), and the charge carrier effective mass change between 1TL, 2TL and 3TL of GaSe. Our ARPES results show that 1TL of GaSe/graphene heterostructures preserve the linear dispersion of graphene and shift the Dirac crossing toward lower binding energy. Furthermore, these results confirm that the band gap can be tuned by controlling the number of GaSe TLs and that the doping level of GaSe can be changed in the proximity of graphene due to the interfacial electron transfer from graphene. The current work provides a basis for further investigation on this 2D material.

## II. METHODS

Few GaSe TL were grown on bilayer graphene/SiC(0001) in a MBE with the base pressure of $5 \times 10^{-10}$ mbar. Elemental Gallium (Ga) of purity of 99.9% and Selenium (Se) of 99.9% in purity were used as the sources, and their fluxes were generated from Knudsen cell. The substrate temperature during film deposition was fixed to about 350 °C. A typical growth rate of about 1.5 nm/min was used; this value was determined based on scanning transmission electron microscopy (STEM). By varying the deposition time, we controlled the number of GaSe layers. The growth was monitored *in situ* by the reflection high-energy electron diffraction (RHEED), which showed streaky patterns signifying two-dimensional growth mode of GaSe. Our four samples were capped with Se to protect them from oxidation. This capping was then removed to perform our STM/STS and ARPES measurements as detailed in the supplementary information (SI) (Figure S1). STM/STS measurements were carried out using an Omicron ultra-high vacuum low temperature scanning tunnelling microscope (UHV-LT-STM). STM/STS were acquired at 77 K in the constant current mode for different bias voltages V, applied to the sample. For the STS measurements, the I(V) characteristics were acquired while the feedback loop was inactive, the differential conductivity dI/dV (V, x, y), proportional to the LDOS, was measured directly by using a lock-in technique. For this purpose a small AC modulation voltage $V_{mod}$ was added to $V$ ($V_{mod,p-p}$= 10 mV, $f_{mod}$= 973 Hz) and the signal $dI$ detected by the lock-in amplifier was used to determine the differential conductivity $dI/dV_{mod}$. The XPS/ARPES experiments were carried out on the TEMPO beamline (SOLEIL French synchrotron facility) at room temperature. The photon energy was selected using a high-resolution plane grating monochromator, with a resolving power E/ΔE that can reach 15,000 on the whole energy range (45 - 1500 eV). During the XPS measurements, the photoelectrons were detected at 0° from the sample surface normal $\vec{n}$ and at 46° from the polarization vector $\vec{E}$. The spot size was 100 × 80 (H×V) μm. The ARPES measurements were carried with hv = 60 eV and a hemispherical electron analyzer with vertical slits to allow band mapping.

## III. RESULTS AND DISCUSSIONS

The GaSe crystal structure is generated from stacking the fundamental building block shown in Figure 1(a). Each building block consists of four covalently bonded Se−Ga−Ga−Se atoms with $D_{3h}$ symmetry and has a lattice constant of 0.37 nm[23] forming a tetralayer (TL). Among the all possible substrate for large and controllable GaSe growth, epitaxial graphene provides several potential advantages for the creation of vdW heterostructure such as: high quality on a wafer scale, high electron mobility, and crystalline ordering that can template commensurate substrate[24,25]. Single and few TL of GaSe were directly grown on bilayer graphene/SiC(0001) substrate with MBE to avoid contamination introduced by chemical transfer[26]. The GaSe films were grown in a MBE equipped with standard effusion cells. During the films growth the substrate was kept at 350 °C, under Se-rich conditions. High-resolution X-ray photoemission spectroscopy (HR-XPS) measurements for the GaSe/graphene heterostructure used in our studies reveal the peaks corresponding to C *1s*, Si *2s*, *2p*, Ga 3*s*, 3*p*, 3*d* and Se 3*s*, 3*p*, 3*d* as shown in the SI (Figure S2). We have prepared different samples for the multiple characterization experiments. The first sample, where about 1.5 layers of GaSe was grown, was used for STM/STS in order to study the tunable band gap within the same sample. The remaining three samples, with 1TL, 2TL, and 3TL of GaSe/graphene, were dedicated to ARPES measurements. It is worth to mention that all of these samples were capped with Se for the sake of stability. A de-capping treatment was needed before starting the measurements, as explained in the SI (Figure S1).

We performed STM/STS measurements on GaSe to simultaneously determine the electronic structure in occupied and unoccupied regimes as well as its spatial dependence[27,28]. Figure 1(b) shows the STM image of a single TL GaSe grown on bilayer graphene. In this area of the sample, the growth was not sufficient to completely cover graphene appearing in the darker region. The height profile in the inset in Figure 1(b) is obtained along the path indicated by the white line. It shows a height change of about 0.8 nm which is in good agreement with the expected thickness of a TL GaSe[29,30]. The spatially resolved differential conductance ($dI/dV$) spectroscopy versus bias voltage spectra, proportional to the LDOS, measured in the dark region of Figure 1(b) is shown in Figure 1(c). The incomplete coverage of the graphene substrate is clearly identified by the well-known V-shaped spectrum (the positions of the Fermi level and the Dirac crossing are highlighted by arrows in the $dI/dV$ curve; the Dirac crossing of Bernal bilayer graphene is shifted of about (–0.35 eV)[31] with respect to the Fermi energy). A completely different U-shaped spectrum in Figure 1(d) indicates the presence of a semiconducting material. This spectrum was taken in the region marked with a red dot in Figure 1(b) *i.e.* on 1TL GaSe terraces to avoid the edges states effect. Figure 1(d) shows that the VBM of 1TL GaSe is located at 2.50 ± 0.05 eV below the Fermi level, and the conduction band minimum (CBM) is located at 1.00 ± 0.05 eV above the Fermi Level, thereby yielding an intrinsic electron quasiparticle bandgap of 3.5 ± 0.05 eV. In contrast to the measured electronic band gap, a significantly smaller optical gap was determined by cathodo-luminescence (CL)[32] measurements which indicated an optical bandgap of 3.3 eV for 1 TL of GaSe. Therefore, an exciton binding energy, namely a difference between the electronic and optical band gap, of 0.2 eV is obtained for 1TL GaSe. This value is smaller than the 0.55 eV determined by Ugeda et al[33] for monolayer $MoSe_2$ on bilayer graphene. This excitonic binding energy could be attributed to the reduced dielectric screening, enhanced Coulomb interactions, and their relatively large effective masses of charge carriers as observed for several 2D $MX_2$ materials.[34–36]

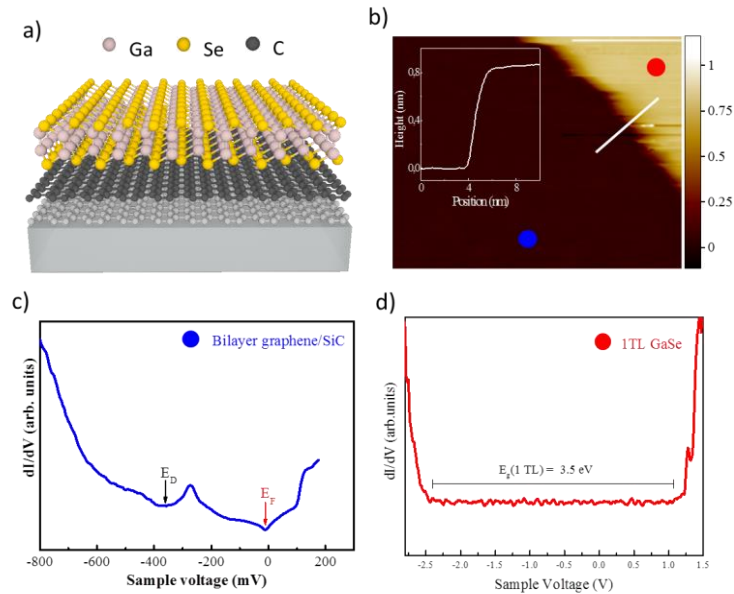

**Figure 1:** a) Sketch of 1TL GaSe on epitaxial graphene, b) Typical STM image (20 nm x 10 nm) of sub-single layer GaSe/graphene ($V_{Tip}$ = 300 meV, $I_t$ = 5 nA, T = 77 K) (Inset: Height profile along the white line reveals a height change of about 8 Å), c) STM d$I$/d$V$ spectrum acquired on bilayer graphene, d) d$I$/d$V$ spectrum showing the electronic bandgap of 1TL GaSe.

The impact of thicker layers on GaSe band gap and the enhanced band gap in a TL GaSe are presented in Figure 2. Figure 2(a) shows a STM topographic image of GaSe/graphene heterostructure where we can differentiate up to 3 TL of GaSe. In addition to the 1TL GaSe over large surface areas, we also obtained two and three TL large islands distributed over the sample surface. The height profile along the red solid line is shown in Figure 2(b). The average apparent height of the three GaSe levels is 8, 16 and 24 Å. The thickness dependence of the electronic band gap is displayed in Figure 2(c). Based on these dI/dV curves, we manage to determine the different gaps[37] of few layered GaSe: the intrinsic 3.50 ± 0.05 eV band gap for a 1TL GaSe is reduced to 3.00 ± 0.05 eV for 2TL and further reduced to 2.30 ± 0.05 eV for 3TL. The uncertainty in the gap values is mainly due to the noise and tip-induced band bending.[34,38,39] We note that the VBM below the FL is shifted from -2.40 ± 0.05 eV for 1TL to -1.30 ± 0.05 eV for 3TL, while the CBM is located at around 1 eV above the Femi level (V = 0). Consequently, the GaSe quasiparticle band gap decreases as the number of TL GaSe increases, inducing also a shift in the VBM position.

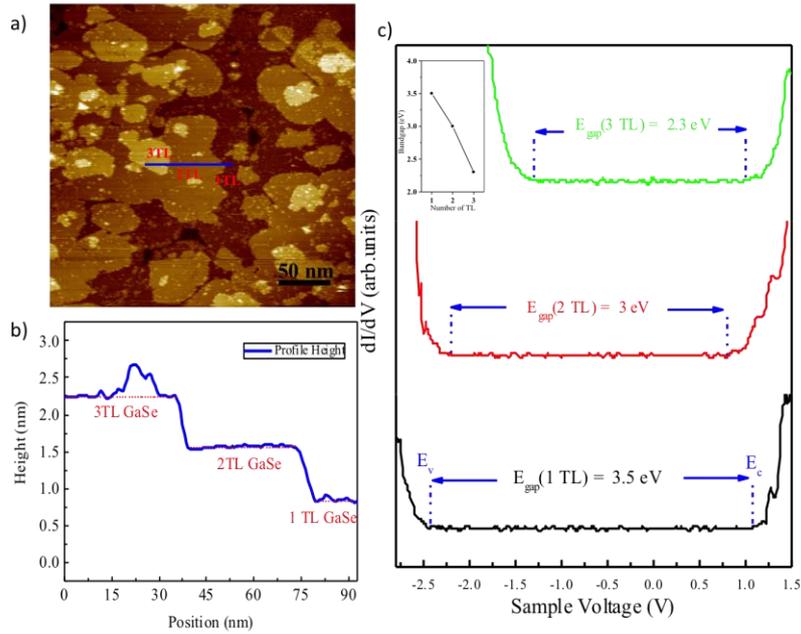

**Figure 2:** Tunable band gap of few TL GaSe/Graphene heterostructure: a) Large-scale (300 nm x 300 nm) STM image shows a GaSe flake containing 1TL, bilayer 2TL and 3TL thickness ($V_{Tip}$= 2.4 V, $I_t$= 5 nA, T= 77 K). b) Lateral profile corresponding to the blue line in panel a), revealing about 8 Å height for each GaSe TL. c) d$I$/d$V$ spectra taken at 1TL, 2TL and 3TL GaSe, respectively, reveals that the bandgap decrease with the increasing thickness (set point: $V_{Tip}$= 1.5 V, $I_{Tip}$= 80 pA); the blue dashed lines correspond to the valence ($E_v$) and the conduction ($E_c$) bands. The inset shows the variation of GaSe bandgap as a function of layer number.

We now discuss the band structure of few layers GaSe/graphene heterostructure. For this, we made three new samples where the thicknesses of GaSe films were controlled so that we get 1TL, 2TL, and 3TL of GaSe. Our ARPES data for 1TL, 2TL, and 3TL GaSe measured at hν= 60 eV along the graphene ΓK direction are shown in Figure 3(a)-(c) respectively. In order to enhance fine spectral features and get better clarity of the band structure presented in Figure 3(a)-(c), the second derivatives of the photoelectron intensity as a function of energy and k-momentum are presented respectively in Figure 3(d)-(f). The measured band structures were compared to our theoretical calculations based on the exchange-correlation potential of the Perdew-Burke-Ernzerhof (PBE) version of the generalized-gradient approximation (GGA)[40]. These calculations, along the GaSe ΓK direction, are in agreement with the ARPES measurements except for some minor details. These slight discrepancies are due to the presence of two GaSe domains: 0° and 30° [41] which can result in the superimposition of the band structure along the ΓM direction in ARPES spectra. From the ARPES maps in Figure 3(a) and (d) corresponding to 1TL of GaSe, one can clearly notice the Mexican hat shaped energy dispersion as a typical signature of single layered GaSe (see inset). This particular shape results in a high density of states (DOS) as shown in Figure S3 *i.e.* van Hove singularity near the VBM.[16] We also can note for 1TL GaSe sample the presence of an additional band (at about -5 eV) corresponding to the sigma band of graphene. Because of the screening effect, this band becomes invisible when increasing the GaSe thickness to 2TL and 3TL. For 1TL, 2TL and 3 TL, we can see that the VBM is not located at the Γ point in contrast to GaSe bulk[41] due to the quantum confinement effect. The VBM corresponds to a binding energy of ~ -2.3 ± 0.05 eV, -1.8 ± 0.05 eV and -1.3± 0.05 eV for 1TL, 2TL and 3 TL respectively. These values of the VBM are similar to the position of the valence band found by STS measurements. Additionally, we note the presence of only one upper band for the 1TL, which evolves into two bands in the 2TL and to three bands in 3TL. This detailed evolution of the valence band structure for different GaSe thickness is highly useful in terms of offering a direct way to determine the GaSe TL number and also shows that we monitor perfectly the growth of layered GaSe thin films.

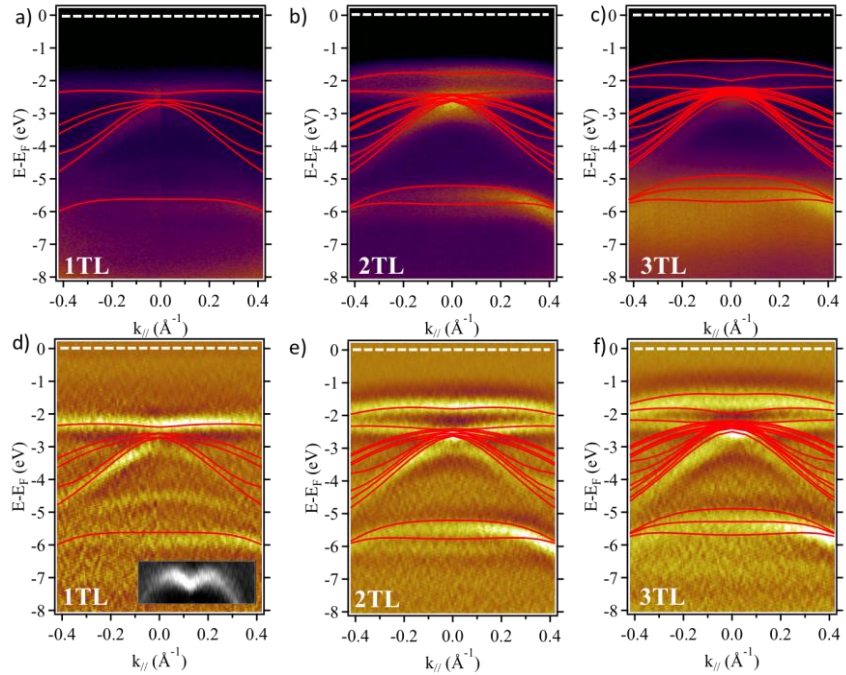

**Figure 3:** Band evolution of few TL GaSe with increasing the GaSe thickness: a), b), and c) ARPES measurements of 1TL, 2TL, and 3TL GaSe/graphene, respectively, at hν= 60 eV along the ΓK direction of GaSe.

d)-f) the second derivative of a)-c) respectively, exhibiting a better visibility of the bands. The Fermi level position is located at the zero of the binding energy (marked in white dotted line). The inset in d): A zoom of 1TL GaSe valence band upper band.

From the ARPES data and the theoretical calculations, the charge carrier effective mass at VBM was determined. The experimental dispersions for 1TL, 2TL, 3TL have been fitted, as in the previous work[41], with a parabolic model $E(k) = E_0 + \frac{\hbar^2}{2m^*} k^2$ where m* is the effective electron mass and ℏ is the reduced Planck constant. We found that the effective mass at VBM is equal to $m^*/m_0$ = -2.16, -1.33, and -1.17 (at $k_{//}$ = 0.26 Å$^{-1}$, 0.16 Å$^{-1}$, and 0.12 Å$^{-1}$) for 1TL, 2TL, and 3TL GaSe, respectively: the charge carrier effective mass decreases by increasing the TL number of GaSe. This reflects the evolution of the electronic structure as varying the thickness of GaSe layers. It is also worth mentioning that the calculated effective mass of 1TL is nearly two times the mass obtained for bulk GaSe at Γ along the ΓK direction[41]; we approach the effective mass of bulk when we increase the GaSe thickness to 3TL.

The ARPES data are in a good agreement with the STM/STS findings except for the slight differences in VBM values of 2TL GaSe. This can be explained by the fact that the STM/STS is a local characterization method while the ARPES scans are averaged on a surface of the order of few hundreds of μm$^2$. Considering a band gap of about 3.5 eV for 1TL and 3 eV for 2TL, as calculated from STM/STS, and the position of the Fermi level with respect to the top of the valence band, from ARPES and STM/STS results, we can infer that the as-grown single and bi TL GaSe are n doped. The position of the Fermi level inside the gap indicates that GaSe becomes less n doped as the number of TL increases. This n doping is clearly obvious for 1TL and it becomes less pronounced for 3TL since this doping effect is lowered when getting further away from the interface. The origin of this n doping could be attributed to an electron transfer from n-doped bilayer graphene to GaSe layer similarly to what was reported for GaSe/graphene heterostructure[42,43]. In other words, the GaSe becomes more p doped, as expected for a bulk GaSe[4,11,17,44], because of the screening effect caused by the presence of additional layers. Similar phenomena was observed for multilayered graphene on SiC[28].

The charge transfer between GaSe and bilayer graphene was further studied using ARPES[45]. The band structure of pristine bilayer graphene[46] and 1TL GaSe/bilayer graphene heterostructure, collected at the K point along the ΓK direction, are shown in Figure 4(a) and (b), respectively. By increasing the number of GaSe TLs, the electronic structure of graphene becomes less visible due to the screening effect. Thus, for 2TL and 3TL of GaSe/graphene, it is not possible to identify precisely the position of the Dirac crossing. For the pristine graphene, the Dirac crossing of the π band is located around 0.35 eV below the Fermi level. The π bands of graphene determine a Fermi velocity $v_F \sim 1.1 \times 10^6$ m/s, and n-doping layers. The double and robust bands at the K high symmetry points confirm that the graphene bilayer at the GaSe/graphene heterostructure preserved the electronic properties of the pristine graphene close to the Fermi level[46] and that the heterostructure is vdW. The profiles shown in Figure 4(c), corresponding to vertical sections at $k_{//}$= 0 Å$^{-1}$ of ARPES maps, were extracted from Figure 4(a) and (b). These profiles (in Figure 4(c)) evidence that for GaSe/bilayer graphene heterostructure the Dirac crossing is shifted by 100 meV toward higher binding energies with respect to the π-band of pristine graphene[45]. This decrease of the n-type doping in graphene can be explained as a result of the charge transfer from graphene to the GaSe layer[46]. The important shift between the pristine and single TL

GaSe/graphene heterostructure is due to the band alignment at the interface between the two 2D materials, in agreement with STS results. It is also worth to note that no chemical bonding between graphene and GaSe happens as no signature of additional bonds appears in the XPS peaks of C 1s of few layered GaSe on graphene shown in Figure S4. These peaks remain unchanged after the GaSe growth, which confirms again the presence of a vdW interface.

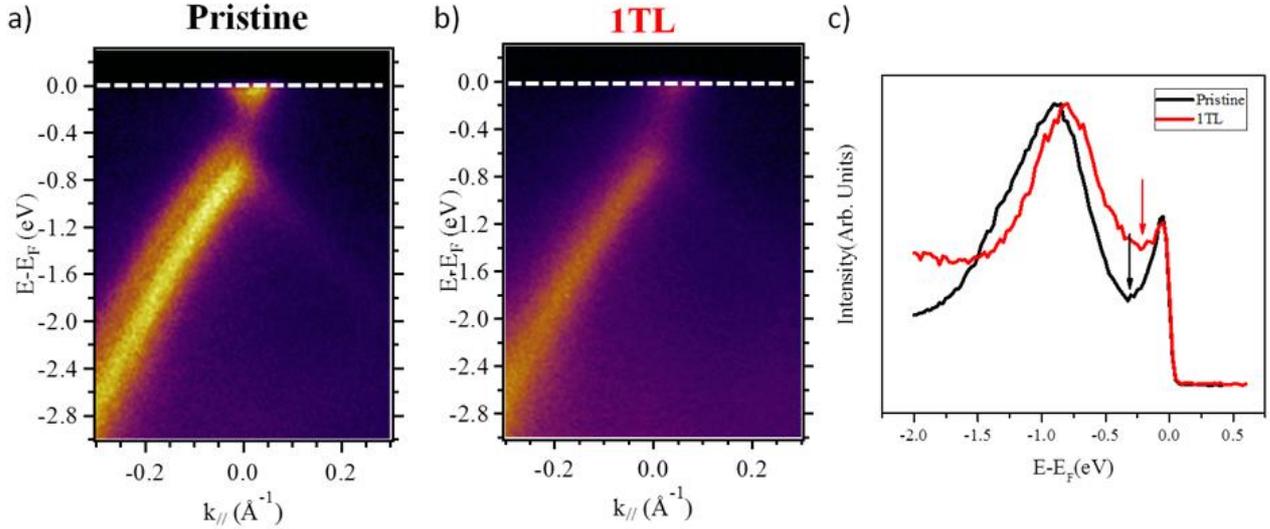

**Figure 4**: Charge transfer between bilayer graphene and 1TL GaSe: a) and b) ARPES at room temperature of pristine graphene and GaSe/graphene heterostructure, measured at hν= 60 eV, through the K-point, along to the graphene ΓK direction; c) ARPES intensity integrated spectra as a function of the binding energy, extracted from the 2D ARPES map, for the initial pristine graphene (black line) and GaSe/graphene (red line). The arrows indicate the positions the Dirac crossing before and after the GaSe growth.

We now briefly discuss the decrease of the GaSe band gap as a function of the number of layers suggesting the tunability of the band gap of ultrathin GaSe layers. Our STS results show the increase of the GaSe bandgap as decreasing the film thickness. Similar results were found with first-principles density functional theory (DFT) calculations using GW approximation[16,47] as illustrated in Table 1. The results obtained by STM/STS and ARPES are in a good accordance with our DFT calculations for 1TL, 2TL, and 3TL of GaSe. This evolution of the band gap, *i.e.* the fact that the band gap decreases when the thickness of MX increases, has been attributed to the changes in the quantum confinement when the number of layers increases as predicted by the DFT calculations[47]. Indeed, the theory has also predicted direct to indirect bandgap transition with decreasing the number of TL of GaSe. It contributes to the decreasing photoluminescence and Raman intensity due to the suppressed inter-layer electron orbital coupling[15,42,48]. These results show how 2D GaSe properties are different from the widely studied $MoS_2$, which has an indirect to direct band gap transition[24] by decreasing the $MoS_2$ layer number. Meanwhile, the energy difference between the gap of a 1TL and few TL GaSe is as large as the energy variation observed when varying the number of layers of $MoS_2$[34]. We can also claim that the VBM and hence the doping level of GaSe are strongly influenced by the electron transfer from n type bilayer graphene to GaSe layers.

|  | 1 TL | 2 TL | 3 TL |
|---|---|---|---|
| **STM/STS** | | | |
| $E_{gap}$ (STS) | 3.5 | 3 | 2.3 |
| VBM | -2.4 | -2.2 | -1.3 |
| CBM | 1.1 | 0.8 | 1 |
| **ARPES** | | | |
| VBM | -2.3 | -1.8 | -1.3 |
| $E_{gap}$ (GW) D. V. Rybkovskiy et al. | 3.9 | 3.1 | 2.8 |
| $E_{gap}$ (GW) T. Cao et al. | 3.7 | - | - |

**Table 1:** Comparison between the measured data using STM/STS and ARPES with the theoretical values reported in literature.

## IV. CONCLUSIONS

In summary, we have successfully studied the electronic bandgap of different thicknesses of GaSe grown on bilayer graphene using MBE. The strong dependency of GaSe bandgap on the number of TLs was demonstrated experimentally by STS. The electronic bandgap decreases from $3.50 \pm 0.05$ eV for 1TL GaSe to $2.30 \pm 0.05$ eV for 3TL GaSe due to the quantum confinement effects. We have also conducted detailed ARPES and theoretical studies of the electronic structure of single, bi and tri TL of GaSe. Contrarily to the p type character of the bulk GaSe, the few-layered GaSe/graphene was shown to be n doped as observed by STS and ARPES. This n type doping is the result of the charge transfer at the interface of this vdW heterostructure. Moreover, this effect is evidenced by the 100 meV shift of the graphene Dirac crossing toward lower binding energies compared to pristine graphene. These results are of great importance for the potential applications of the vdW heterostructures based on layered GaSe.

## ACKNOWLEDGEMENTS

This work was supported by ANR H2DH grants. M. Y. and K. X. acknowledge the Center for Nanophase Materials Sciences, which is a DOE Office of Science User Facility.

# Supplementary Information

# Tunable Quasiparticle Band Gap in Few Layers GaSe/graphene Van der Waals Heterostructures

Zeineb Ben Aziza[1], Debora Pierucci[1], Hugo Henck[1], Mathieu G. Silly[2], Christophe David[1], Mina Yoon[3, 4], Fausto Sirotti[2], Kai Xiao[3], Mahmoud Eddrief[5, 6], Jean-Christophe Girard[1], and Abdelkarim Ouerghi[1, *]

[1] Centre de Nanosciences et de Nanotechnologies, CNRS, Univ. Paris-Sud, Université Paris-Saclay, C2N – Marcoussis, 91460 Marcoussis, France
[2] Synchrotron-SOLEIL, Saint-Aubin, BP48, F91192 Gif sur Yvette Cedex, France
[3] Center for Nanophase Materials Sciences, Oak Ridge National Laboratory, Oak Ridge, Tennessee 37831, USA
[4] The University of Tennessee, Knoxville, Tennessee 37996, USA
[5] Sorbonne Universités, UPMC Univ. Paris 06, UMR 7588, INSP, F-75005 Paris, France
[6] CNRS, UMR 7588, Institut des NanoSciences de Paris (INSP), F-75005 Paris, France

*Corresponding author, E-mail: abdelkarim.ouerghi@lpn.cnrs.fr, fax number: +33169636006

## Capping and de-capping GaSe:

To remove the Se capping layer of GaSe films in UHV chamber before STM/STS and ARPES experiments, a very careful annealing of the sample was performed. Since the GaSe film starts to decompose at 300 °C and the Se capping layer could be removed in a temperature range of 200–280 °C, it is recommended to perform the de-capping procedure at lower temperature; this is quiet challenging for typical temperature measurements. Therefore, we gradually increased the annealing current and simultaneously monitored the samples by Reflection High Energy Electron Diffraction (RHEED). The appearance of a sharp RHEED pattern of the GaSe films (Figure S1) indicates the proper de-capping temperature. Several attempts of annealing procedures were performed to optimize the de-capping temperature.

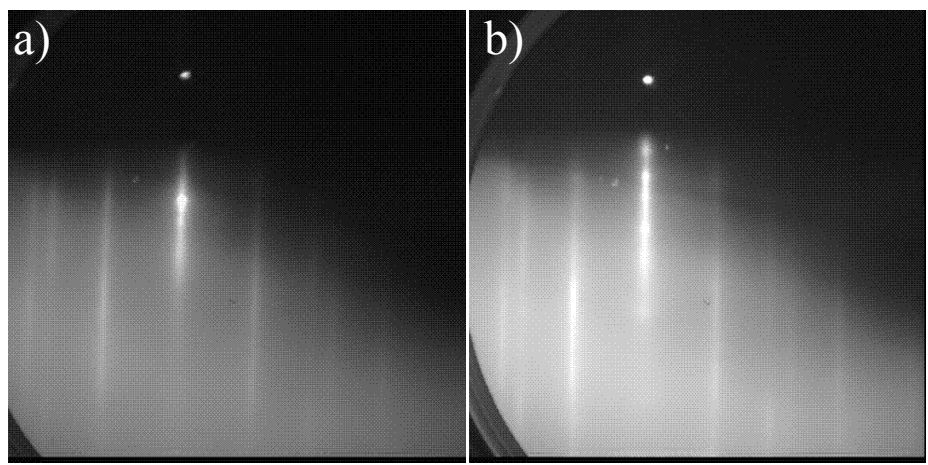

**Figure S1.** The RHEED patterns of GaSe/graphene a) before and b) after the de-capping; the bright contrast in b) indicates the successful de-capping procedure.

**XPS:**

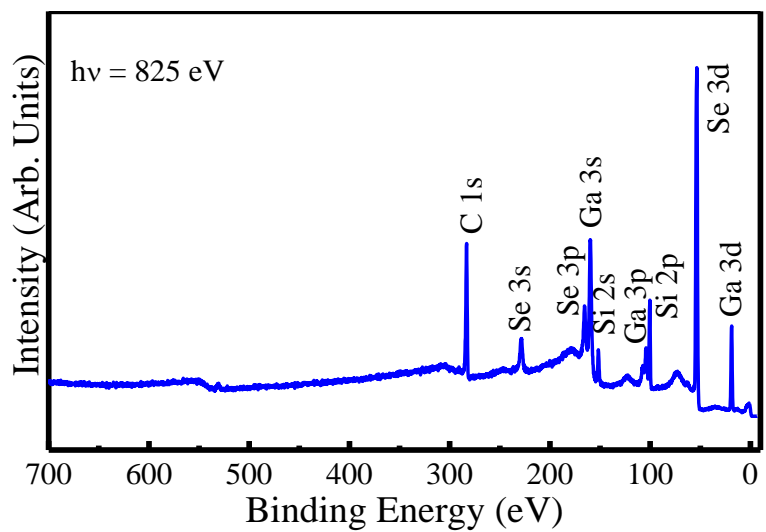

**Figure S2.** XPS spectra of single layered GaSe/graphene heterostructure measured at hν = 825 eV.

**ARPES:**

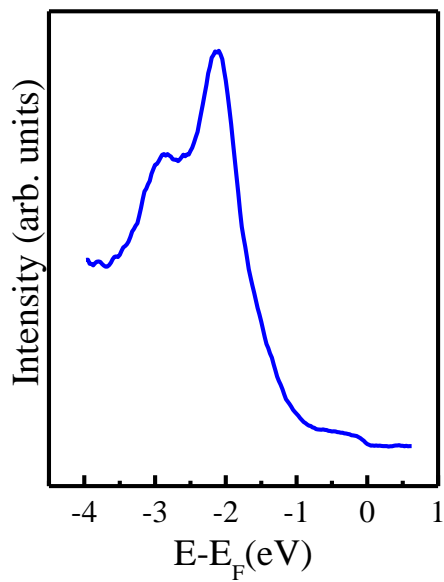

**Figure S3.** ARPES intensity integrated spectrum as a function of the binding energy, extracted from the 2D ARPES map of 1TL GaSe.

**XPS:**

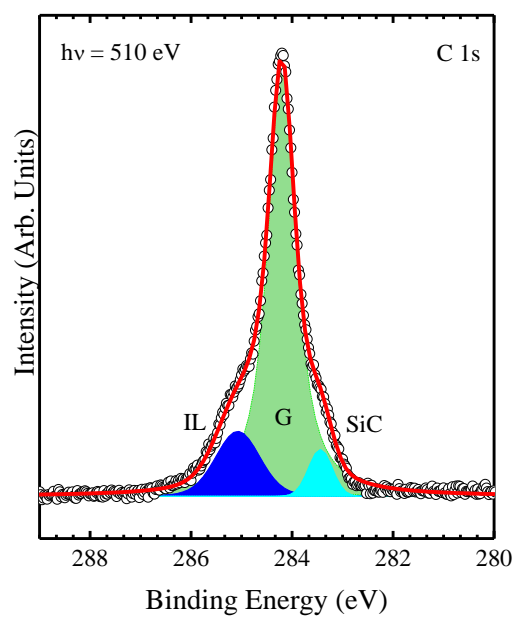

**Figure S4.** XPS peaks of C 1s (measured at hν = 510 eV) of few layered GaSe on graphene. The experimental data points are displayed as dots. The solid line is the envelope of fitted components. (IL: interfacial layer; G: graphene).